\documentclass[5p]{elsarticle}
\usepackage{microtype}
\usepackage[utf8]{inputenc}
\usepackage{amsmath}

\usepackage{amsthm,amsfonts}
\usepackage{nicefrac}
\usepackage{amssymb,upgreek}
\usepackage{hyperref}
\usepackage[UKenglish]{babel}
\usepackage[T1]{fontenc}
\usepackage{csquotes}

\newcommand{\kl}[1]{\left( #1 \right)}
\newcommand{\kle}[1]{\left[ #1 \right]}
\newcommand{\ed}[1]{\frac{1}{#1}}
\newcommand{\defi}{\mathrel{\mathop:}=}
\newcommand{\ifed}{=\mathrel{\mathop:}}
\newcommand{\dif}{\ensuremath{\operatorname{d}}\!}
\newcommand{\Li}[1]{\ensuremath{\operatorname{Li}_{ #1 }}}
\newcommand{\kB}{k_\text{B}}
\begin{document}
	\begin{frontmatter}
		\title{Sparsity of Hawking Radiation in $D+1$ Space-Time Dimensions Including Particle Masses\tnoteref{t1}}
		\tnotetext[t1]{This document contains results of a PhD thesis funded by the Victoria University of Wellington PhD Scholarship.}
		
		\author{Sebastian Schuster\fnref{fn1}}
		\ead{sebastian.schuster@sms.vuw.ac.nz}
		\address{School of Mathematics and Statistics, Victoria University of Wellington, PO Box 600, Wellington 6140, New Zealand}
		\date{\today}
		
		\begin{abstract}
			Hawking radiation from an evaporating black hole has often been compared to black body radiation. However, this comparison misses an important feature of Hawking radiation: Its low density of states. This can be captured in an easy to calculate, heuristic, and semi-analytic measure called \enquote{sparsity}. In this letter we shall present both the concept of sparsities and its application to $D+1$-dimensional Tangherlini black holes and their evaporation. In particular, we shall also publish for the first time sparsity expressions taking into account in closed form effects of non-zero particle mass. We will also see how this comparatively simple method reproduces results of (massless) Hawking radiation in higher dimensions and how different spins contribute to the total radiation in this context. 
		\end{abstract}
	
		\begin{keyword}
			Hawking Radiation \sep Black Holes \sep Higher Dimensions
		\end{keyword}
	\end{frontmatter}
	
	\section{Introduction to Sparsity}
	One of the core features of Hawking radiation \cite{HawkingEffect1,HawkingEffect2} is the similarity of its emission spectrum to that of a grey body, hence that of a black body. While educational and edifying, this comparison omits one very important feature: This radiation is \emph{sparse}. Technically, this concept is encoded in a low density of states \cite{KieferDecoherenceHawkingRad,BHradNonclass}. Strictly speaking and as evidenced by Don Page's work \cite{Page1,Page2,PageThesis,Page3}, this feature is known since the beginning, yet is often glossed over. A subsequent focus on high temperature regimes can be considered partly responsible for this \cite{OliensisHill84,BHEvapJets,NatureBHgamma,BHPhotospheresPage,BremsstrahlungBH,MG12PageMacGibbonsCarr}. Four years ago, a heuristic method to bring this forgotten feature more to the forefront has been introduced \cite{HawkFlux1,HawkFlux2,SparsityNumerical} --- simply called \emph{sparsity} $\eta$. This concept was originally applied to $3+1$ dimensional black holes (corresponding to the solutions of Schwarzschild, Kerr, Reissner--Nordström, and \enquote{dirty black holes}), but soon found application also in higher dimensions \cite{HodNdim}, in phenomenological quantum gravity extensions \cite{SparsityBackreaction}, and in the context of the influence of generalised uncertainty principles on Hawking evaporation \cite{SparsityAna,OngGUPSparsity}.
	
	Let us quickly introduce the concept: Sparsity $\eta$ is a measure to estimate the density of states of radiation. For this, one compares a localisation time scale $\tau_{\text{loc}}$ of an emitted particle with a time scale $\tau_{\text{gap}}$ characterising the time between subsequent emission events. It is worth emphasising the indefinite article here: Different choices can be made for both time-scales, though their numerical values will not differ by much. The easiest way to choose the time scale $\tau_{\text{gap}}$ is given by the inverse of the integrated number flux density $\dif \Upgamma_n$ of the radiation, and we will adhere only to this choice throughout the letter. Hence:
	\begin{equation}
		\tau_{\text{gap}} = \ed{\Upgamma_n}.
	\end{equation}
	For the localisation time scale $\tau_{\text{loc}}$, however, the identification of a \enquote{simplest} choice is less obvious. This is due to the fact that for most spectra (in our context the Planck spectrum) peak and average frequencies do not agree, nor are they the same when comparing number density spectrum $\Upgamma_n$ and energy density spectrum $\Upgamma_E$. We encode this in the following way:
	\begin{equation}
		\tau_{\text{loc}} = \ed{\nu_{c,q,s}} = \frac{2\pi}{\omega_{c,q,s}},
	\end{equation}
	where the index $c \in \{ \text{avg.}, \text{peak}\}$ indicates how the physical \underline{q}uantity $q$ (associated to a unique frequency by appropriate multiplication with natural constants $\hbar,\kB,c,G$) is \underline{c}alculated, and $s$ determines the \underline{s}pectrum we consider\footnote{Only in the definition of $\tau_{\text{loc}}$ we employed the frequency, not the angular frequency as it makes for a more conservative estimate of sparsity. This was suggested by an anonymous referee of \cite{HawkFlux1}.}. As our $\tau_{\text{gap}}$ is fixed, the sparsities
	\begin{equation}\label{eq:defsparsity}
		\eta_{c,q,s} \defi \frac{\tau_{\text{gap}}}{\tau_{\text{loc}}}
	\end{equation} 
	inherit the freedom (\textit{i.e.}, the indices) of $\tau_{\text{loc}}$. The method of calculating the sparsities is always the same: First, calculate the quantity $q$ associated to the spectrum $s$. Second, find a corresponding angular frequency $\omega_{c,q,s}$. Here, the choice of the quantity is of relevance --- one should keep to quantities which can be related in a straightforward manner to a frequency. Third, and last, calculate the sparsity
	\begin{equation}
		\eta_{c,E,s} = \frac{\omega_{c,E,s}}{2\pi \, \Upgamma_n}.
	\end{equation}
	Let us see this in action on two less trivial examples: Here, the frequency is gained from either the average wavelength $\lambda$ or the average period $\tau$ of emitted particles for the number spectrum. In order to compare the frequency corresponding to the average wavelength or of the average period of emitted particles with $1/\Upgamma_n$ as it appears in the sparsity definition~\eqref{eq:defsparsity}, one has to take their inverses before multiplying with the appropriate factors of speed of light $c$, and Planck's constant $\hbar$. In the resulting expression, $\Upgamma_n$ cancels and one arrives at the convenient expressions
	\begin{equation}
		\eta_{\text{avg.},\tau,n} = \ed{\int \frac{2\pi \hbar}{E} \dif \Upgamma_n}, \qquad \eta_{\text{avg.},\lambda,n} = \ed{\int \frac{2\pi}{c k} \dif \Upgamma_n},
	\end{equation}
	where $E$ is the energy of the emitted particle (equivalent to its angular frequency, as $E=\hbar \omega$), and $k$ its wave number. Note that both integrals are identical \emph{in the case of massless particles} --- in this case we simply call both the same, $\eta_{\text{binned}}$. A convenient bonus of this result is the fact that different emission processes can be considered \enquote{happening in parallel}, with the relevant notion borrowed from circuit analysis, that is
	\begin{equation}\label{eq:binning}
		\ed{\eta_{\text{tot}}} = \sum_{\text{channel}~i} \ed{\eta_i}
	\end{equation}
	for these two sparsities (and others reducing to a simple inverse of a single integral). They also lend themselves nicely to a interpretation as \enquote{binned} or \enquote{bolometric} sparsity measures, as they can be understood as dividing up the emission spectrum into infinitesimal bins.
	
	Much more common, however, is an expression not amenable to this property. For example, if one calculates the peak frequency (\emph{i.e.}, the peak energy) of the number spectrum, the resulting sparsity
	\begin{equation}
		\eta_{\text{peak},E,n} = \frac{\omega_{\text{peak},E,n}}{2\pi \Upgamma_n}
	\end{equation}
	involves finding the zeroes of the first derivative of $\Upgamma_n$ w.r.t. energy/frequency. However, the peak frequencies will usually not be given explicitly, as only in rare special cases can they be found analytically exactly. 
	
	This is an opportune moment to describe the spectra under consideration in more detail. We shall consider spectra of the form
	\begin{equation}
		\dif\Upgamma_n = \frac{g}{(2\pi)^D} \frac{c \hat{k}\cdot\hat{n}}{\exp\kl{\frac{\hbar\sqrt{m^2 c^2+k^2 c^4}}{\kB T} - \tilde\mu} + s} \dif^D k \dif A
	\end{equation}
	and
	\begin{equation}
		\dif\Upgamma_E = \frac{g}{(2\pi)^D} \frac{c \hbar\sqrt{m^2 c^2+k^2 c^4} \; \hat{k}\cdot\hat{n}}{\exp\kl{\frac{\hbar\sqrt{m^2 c^2+k^2 c^4}}{\kB T} - \tilde\mu} + s} \dif^D k \dif A,
	\end{equation}
	where $g$ is the (possibly dimension-dependent) degeneracy factor of the emitted particles (more below in section~\ref{sec:massless}), $m$ their mass, $D$ the number of space dimensions, $T$ the temperature of the radiation, $\tilde\mu$ the chemical potential divided by $\kB T$ (\emph{i.e.}, the logarithm of the fugacity), $\hat{n}$ the surface normal to the emitting hypersurface $A$, and $s \in \{-1,0,+1\}$ a parameter distinguishing (respectively) between bosons, Maxwell--Boltzmann/classical particles, and fermions. The differential $\dif^D k$ takes on the following form in spherical coordinates:
	\begin{align}
		&\dif^{D} k\nonumber\\& = k^{D-1} \sin^{D-2} \varphi_1 \cdots \sin \varphi_{D-2} \sin^0 \varphi_{D-1} \dif k \dif \varphi_1 \dif\varphi_{D-1},
	\end{align}
	where $\varphi_{D-1}\in[0,2\pi)$, $\varphi_i \in [0,\pi)$, if $i \in \{2,\dots,D-2\}$, and $\varphi_{1} \in [0,\frac{\pi}{2})$ (at least for our future integration steps). In many cases, the term $\hat{k} \cdot \hat{n}$ seems to be forgotten in higher dimensions (we will not mention the guilty parties) --- even though without it, it will not be possible to correctly link these spectra to the $3+1$ dimensional case and its Stefan--Boltzmann law.
	
	Regarding our earlier mentioned peak frequencies in $3+1$ dimensions, the peak frequencies of the classical massive particle's number and energy spectra can be found in terms of cubics --- but even these are neither useful nor enlightening in most situations. For massless particles, on the other hand, the result is in all dimensions expressible in terms of the Lambert~W-function:
	\begin{subequations}\label{eq:peaks}
		\begin{align}
			&m=0:\nonumber\\
			 &\omega_{\text{peak},E,E} = \frac{\kB T}{\hbar}\kl{D+W(sD e^{\tilde\mu-D+2})},\\
			 & \omega_{\text{peak},E,n} = \frac{\kB T}{\hbar}\kl{D-1+W(s(D-1) e^{\tilde\mu-D+3})}.
		\end{align}
	\end{subequations}
	
	In the following we will apply these methods to Tangherlini black holes. The sparsities we will calculate in this letter are: $\eta_{\text{peak},E,n}$, $\eta_{\text{peak},E,E}$, $\eta_{\text{avg.},E,n}$, $\eta_{\text{avg.},\tau,n}$, and $\eta_{\text{avg.},\lambda,n}$. The latter two being the same in the massless case, they are relabelled as $\eta_{\text{binned}}$ in that case.
	
	\section{Preliminaries for Tangherlini Black Holes}
	
	The Tangherlini black hole \cite{Tangherlini,FrolovZelnikov2011} is the higher dimensional generalization of the Schwarzschild black hole; it is the $D+1$-dimensional, spherically symmetric vacuum solution. The metric has the form
	\begin{align}\label{eq:Tangherlini}
		\dif s^2 =& -\kl{1-\kl{\frac{r_\text{H}}{r}}^{D-2}}\dif t^2 + \kl{1-\kl{\frac{r_\text{H}}{r}}^{D-2}}^{-1}\dif r^2 \nonumber\\&+ r^2 \dif \Omega_{D-1}^2,
	\end{align}
	where $\dif\Omega_{D-1}^2$ is the differential solid angle, and
	\begin{equation}
		r_\text{H} = \sqrt[D-2]{\frac{8 \Gamma(\frac{D}{2})GM/c^2}{(D-1)\pi^{\nicefrac{(D-2)}{2}}}}
	\end{equation}
	is the $D+1$-dimensional Schwarzschild radius, $G$ the (dimension-dependent) gravitational constant, and $M$ the mass of the black hole. A $\Gamma$ without the indices indicating number or energy densities simply refers to the $\Gamma$-function. In passing, we note that the uniqueness theorems for black holes hold (without further assumptions) only in $3+1$ dimensions \cite{Papantonopoulos2009,NumBlackSaturns,QGHolo}, related to a more complex notion of angular momenta. This somewhat justifies our focus on higher dimensional, non-rotating black holes even though some solutions are explicitly known, like the Myers--Perry solution \cite{MyersPerry}.
	
	The surface area of the horizon becomes
	\begin{equation}
		A_{\text{H}} = 2 \frac{\pi^{D/2}}{\Gamma(D/2)} r_{\text{H}}^{D-1},
	\end{equation}
	while the corresponding Hawking temperature is
	\begin{equation}
		\frac{D-2}{4\pi r_\text{H}}\frac{\hbar c}{\kB}.
	\end{equation}
		
	Due to the spherical symmetry, the angular and area integral required for the sparsity calculations can (a) be separated from each other, and (b) the term $\hat{k}\cdot\hat{n}$ evaluates to a simple $\cos\varphi_1$. This factor will prevent an integration over the angular variables from being the area of a hypersphere. Rather, the result is (in all instances to be encountered in the following)
	\begin{align}
		&\int_0^{2\pi} \dif \varphi_{D-1} \int_{0}^{\pi}\dif \varphi_{D-2} \sin \varphi_{D-2} \times\cdots \nonumber\\
		&\times \int_{0}^{\pi}\dif \varphi_2 \sin^{D-3}\varphi_2 \int_{0}^{\frac{\pi}{ 2}} \dif \varphi_1 \cos\varphi_1 \sin^{D-2}\varphi_1\nonumber\\
		&\quad = \frac{2\pi}{D-1}\frac{\sqrt{\pi}^{D-3}}{\Gamma(\ed{2}(D-1))}.
	\end{align}
	
	\section{Sparsity Results and Comparison with the Literature}
	
	The origin of the sparsity of Hawking evaporation in $3+1$ dimensions can be sought and found in the connection between size of the horizon and the Hawking temperature. This feature is absent from black bodies --- as long as their temperature can be maintained, they can be made of arbitrary sizes. Put differently, in $3+1$ dimensions and for non-rotating\footnote{Since, as mentioned before, rotating black hole solutions are more subtle in higher dimensions we will limit the discussion to non-rotating ones. As a shorthand, we will from now on assume no rotation.} black holes, the thermal wavelength $\lambda_{\text{thermal}}$ fulfils $\lambda_{\text{thermal}}^2 < A_{\text{H}}$. However, as we will see below, this does not translate to arbitrary dimensions as already shown by Hod in \cite{HodNdim}. Also, the inclusion of rotation would lead to the emergence of super-radiance further complicating the discussion. However, away from super-radiant regimes one can include easily the parameter $\tilde\mu$ (introduced above) to capture at least charges --- allowing a spherically symmetric solution ---, or with less qualms and more bravado about deviating from spherical symmetry even very small angular momenta.
	
	First, we will reproduce and improve Hod's results on the emission of massless particles, then we shall generalise to massive particles. Due to the length of the results, these will be provided in tables~\ref{tab:massless} and~\ref{tab:massive}. All results will be given in terms of $\nicefrac{\lambda_{\text{thermal}}^{D-1}}{gA}$. Note that in this expression the Tangherlini black hole mass drops out.
	
	\subsection{Warm-up: The Massless Case}\label{sec:massless}
	
	Before starting, it is worth reminding ourselves that we want to be as conservative as possible in our sparsity results: A small sparsity would mean little phenomenological departure from the familiar black body radiation. Hence, we will not consider the area of the horizon to be the relevant area from which the Hawking radiation originates, but rather we will take the capture cross section $\sigma_{\text{capture}}$ for massless particles. This turns out to be \cite{FrolovZelnikov2011}
	\begin{equation}
		\sigma_{\text{capture}} = \underbrace{\ed{2\sqrt{\pi}}\frac{\Gamma(\nicefrac{D}{2})}{\Gamma(\nicefrac{D+1}{2})} \kl{\frac{D}{D-2}}^{\frac{D-1}{2}} \kl{\frac{D}{2}}^{\frac{D-1}{D-2}}}_{\ifed c_\text{eff}} A_\text{H}.
	\end{equation}
	The factor $c_\text{eff}$ has been defined for future convenience. This change of area can be motivated and backed with numerical studies highlighting that the renormalised stress-energy tensors of Hawking radiation do not have their maximum at or very close to the horizon but rather a good distance away from it \cite{BHQuantumAtmosphere}.
	
	It is relatively straightforward (though not necessarily notationally easy-going) to manipulate standard integral expressions \cite{GradshteynRyzhik1980,Olver2010a} into the required form. The Boltzmann case (\emph{i.e.}, $s=0$) often requires recognising removable singularities, but apart from this is straightforward to include in these results. This is most apparent in the ubiquitous expressions $\Li{n}(-s)/(-s)$ involving the polylogarithm of order $n$. We have collected the results in table~\ref{tab:massless}. In order to emphasise the dependence of the degeneracy factor $g$ on the dimension, it is written as $g(D)$ in the table.
	
	\begin{table*}
		\centering
		\begin{tabular}{rcl}
			$\eta_{\text{peak},E,n}$ &$=$& $\displaystyle\frac{1}{2\pi (D-2)!} \frac{\Gamma(\frac{D-1}{2})}{\pi^{\nicefrac{(D-3)}{2}}}\frac{(D-1+W((D-1)s e^{\mu-D+3}))}{\frac{\Li{D}(-s e^{\mu})}{(-s)}} \frac{\lambda_\text{thermal}^{D-1}}{g(D) c_\text{eff} A_\text{H}}$\\
			$\eta_{\text{peak},E,E}$ &$=$& $\displaystyle\frac{1}{2\pi (D-2)!} \frac{\Gamma(\frac{D-1}{2})}{\pi^{\nicefrac{(D-3)}{2}}}\frac{(D+W(D s e^{\mu-D+2}))}{\frac{\Li{D}(-s e^{\mu})}{(-s)}} \frac{\lambda_\text{thermal}^{D-1}}{g(D) c_\text{eff} A_\text{H}}$\\
			$\eta_{\text{avg.},E,n}$ &$=$& $\displaystyle\frac{D}{2\pi (D-2)!} \frac{\Gamma(\frac{D-1}{2})}{\pi^{\nicefrac{(D-3)}{2}}}\frac{\frac{\Li{D+1}(-s e^{\mu})}{(-s)}}{\kl{\frac{\Li{D}(-s e^{\mu})}{(-s)}}^2} \frac{\lambda_\text{thermal}^{D-1}}{g(D) c_\text{eff} A_\text{H}}$\\
			$\eta_{\text{binned}}$ &$=$& $\displaystyle\frac{\Gamma(\nicefrac{(D-1)}{2}) (D-1)}{2\pi\sqrt{\pi}^{D-3} (D-2)!}\frac{1}{\frac{\Li{D-1}(-se^{\mu})}{(-s)}} \frac{\lambda_\text{thermal}^{D-1}}{g(D) c_\text{eff} A_\text{H}}$
		\end{tabular}
		\caption{Sparsities for emission of massless particles in a $D+1$-dimensional Tangherlini space-time in terms of polylogarithms $\Li{n}(x)$, and Lambert-W functions $W(x)$. $\lambda_{\text{thermal}}$ is the thermal wavelength, $c_\text{eff}$ a correction factor to link capture cross-section with horizon area $A_\text{H}$, and $g(D)$ the particles' degeneracy factor.}
		\label{tab:massless}
	\end{table*}
	
\begin{figure}
		\centering
		\includegraphics[width=\columnwidth]{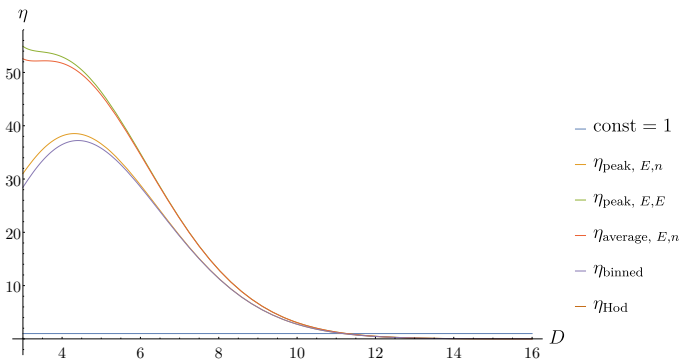}
		\caption[Plot of Dependence of Sparsity of Gravitons in Dimension D]{A comparison of various sparsities $\eta$ for massless gravitons in $D$ space dimensions. The constant $1$ indicates the transition sparse to non-sparse.}
		\label{fig:ndim}
	\end{figure}

	These results correctly reproduce the earlier, $3+1$-dimensional results found in \cite{HawkFlux1,MyThesis}. Note that the exact solution of the peak frequencies of equations~\eqref{eq:peaks} has a different asymptotic behaviour for $D\to \infty$ compared to the approximation used in \cite{HodNdim}. This does not influence the general statement much: Sparsity is lost in high dimensions, Hawking radiation indeed becomes classical and fully comparable to a black body spectrum. However, the exact dimension where the transition sparse to non-sparse happens changes. In figure~\ref{fig:ndim} we compare the various sparsities and their dependence on $D$ for massless gravitons as done in \cite{HodNdim}, where $\eta_{\text{Hod}} \approx \frac{e}{8\pi^2}\kl{\frac{4\pi}{D}}^{D+1}$. We can see that the qualitative picture each measure of sparsity draws is universal --- and at least in the massless case this can be inferred from the way numerator and denominator behave in the definition~\eqref{eq:defsparsity}.
	
	This is especially important once one takes into account the fact that we ignore grey body factors throughout our calculation: Their inclusion will push non-sparsity necessarily to even higher dimensions. That even their inclusion will not change the qualitative result, is nonetheless shown by another comparison to the literature: In \cite{KantiNdimBH,CardosoBulkHawking,KantiNdimBH2} numerical analysis was performed to take the effects of grey body factors into account. Even though these analyses were performed without sparsity as such in mind, it is easy to compare how different particle types will behave. As the different degeneracy factors $g$ for massless particles with different spin depend characteristically on the space dimension $D$, let us summarise these for the spins present in the standard model of particle physics plus gravity: $g_\text{scalar} = 1$, $g_{\text{spin~}1/2} = 2^{n-1}$ for $D=2n$ or $D=2n-1$ and assuming Dirac fermions (and counting particles and anti-particles separately), $g_\text{vector} = D-1$, and $g_\text{graviton} = (D+1)(D-2)/2$. These degeneracies depend, however, on the specifics of the higher dimensional physics considered: In brane world models they are for all $D$ the familiar, $3+1$-dimensional ones for emission into the brane \cite{KantiNdimBH2}. Such brane world models are already covered by the present analysis --- up to a dimension-\emph{independent} factor this corresponds to looking at the dimension-dependence of the scalar sparsities.
	
\begin{figure}
		\includegraphics[width=\columnwidth]{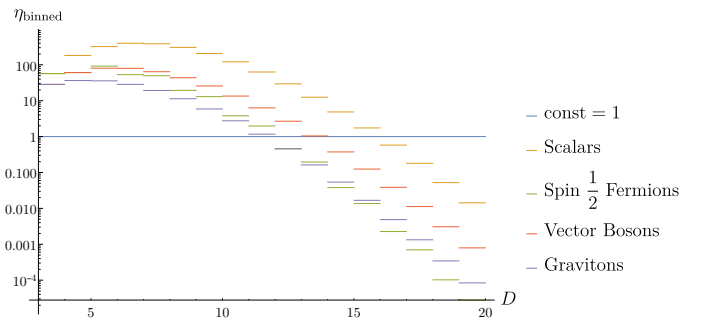}
		\caption[Plot of Dependence of Binned Sparsity of Different Particles in Dimension D]{The binned sparsity $\eta_{\text{binned}}$ for different (massless) particle species.}
		\label{fig:species}
	\end{figure}

	The $3+1$-dimensional case shows amply \cite{HawkFlux1,SparsityNumerical} that the inclusion of grey body factors drastically changes the sparsity of, for example, gravitons. Even so, as shown in figure~\ref{fig:species}, the simplifications made while deriving our expressions for sparsity still qualitatively reproduce the behaviour of the earlier-mentioned, numerical studies. While the order in which different particles change from $\eta>1$ to $\eta<1$ shows minor changes, the over-all behaviour is retained, as is the prediction that emitted gravitons become classical radiation first. This then would correspond to a thermal gravitational wave.
	
	\subsection{Gory Details: The Massive Case}\label{sec:massive}
	Before starting the calculations for massive particles, it is a good idea to revisit the effective area $A=c_\text{eff} A_\text{H}$. The capture cross-section underlying this approach changes significantly for massive particles: They become dependent on the particle's velocity $\beta$. While for any massless particle $\beta=c$, for massive particles this means that the effective capture cross-section diverges to $\infty$ for particles with velocity $\beta=0$. The capture cross-section for massless particles reappears as the limiting case for $\beta\to 1$. Finding a corresponding effective cross-section for any given $\beta$ can still be done analytically in $3+1$ dimensions, but this fails in higher dimensions. On top of this, in higher dimensions stable orbits do not exists \cite[\S7.10.2]{FrolovZelnikov2011}; at least assuming the dynamics of higher dimensional general relativity. 
	
	To retain an ansatz for the following calculation we shall hence assume that the same effective cross-sectional area as for massless particles gives a good approximation for the area from which Hawking radiation originates. On the one hand, in $3+1$ dimensions this seems a good starting point as we can expect the massless case to be a limiting case for massive particles. An example of this approach is found in \cite{BHQuantumAtmosphere}: The heuristic arguments based on an analogy to the Schwinger effect presented therein cover both massive and massless cases; the additionally studied renormalised stress-energy tensor for massless particles constitutes such a limiting case. On the other hand, the assumption that this carries over in some way to higher dimensions is also a good starting point and working hypothesis.
	
	These arguments in place, we can head straight for the integrals involved, only this time with the relation $E^2 = k^2c^2 + m^2c^4$ connecting the momentum $k$ and the energy $E$ of the particle emitted. The strategy here is always similar: First simplify the integration by rewriting it as the integration of a geometric sum, then integrating by parts until one can make use of the substitution $k=z\cosh x$. This allows employing the identity \cite[3.547.9]{GradshteynRyzhik1980}:
		\begin{gather}
			\int_{0}^\infty \exp\kl{-\beta\cosh x} \sinh^{2\nu} x\dif x = \hspace{2cm} \nonumber \\\ed{\sqrt{\pi}}\kl{\frac{2}{\beta}}^\nu \Gamma\kl{\frac{2\nu+1}{2}}K_\nu(\beta),
		\end{gather}
	valid for $\mathrm{Re}(\beta)>0, \mathrm{Re}(\nu) >-1/2$. The resulting sums of modified Bessel functions of the second kind are the expressions in table~\ref{tab:massive}.
	
	At first glance, these seem to be rather unhelpful for further analysis. This is not quite the case: For example, remembering that for fixed $\nu$
		\begin{equation}
			K_\nu(\beta) \stackrel{\beta\to \infty}{\sim} \sqrt{\frac{\pi}{2\beta}} e^{-\beta},
		\end{equation}
	tells that for high masses sparsity will be regained in any (fixed) dimension. From a phase-space point of view this is what physical intuition would suggest. Likewise, asymptotic expansions for $z\to 0$ will regain our earlier, massless results. Similar asymptotic analysis was employed in the service of separating superradiant regimes from genuine Hawking radiation in the analysis of the Kerr space-time in \cite{MyThesis} and \cite{HawkFlux1} (though it involved modified Bessel functions of the \emph{first} kind and requires restricting oneself to sparsities fulfilling property~\eqref{eq:binning}).
		
		\begin{table*}
			\centering
			\resizebox{\textwidth}{!}{
			\begin{tabular}{rcl}
				$\eta_{\text{peak},E,n/E}$ &$=$& $\displaystyle \frac{(D-1)}{\sqrt{\pi}^{D-2} 2^{\nicefrac{D+3}{2}}}\frac{\Gamma\kl{\frac{D-1}{2}}}{\Gamma\kl{\frac{D+2}{2}}}\frac{\omega_{\text{peak},E,n/E}}{z^{\frac{D+1}{2}}} \kle{\displaystyle \sum_{n=0}^{\infty} \frac{(-s)^n e^{(n+1)\tilde\mu}}{(n+1)^{\frac{D-1}{2}}} K_{\nicefrac{D+1}{2}}\kl{(n+1)z}}^{-1} \frac{\lambda_\text{thermal}^{D-1}}{g(D) c_\text{eff} A_\text{H}}$\\
				$\eta_{\text{avg.},E,n}$ &$=$& $\displaystyle \frac{D(D-1)}{2^{\nicefrac{(D+3)}{2}} \sqrt{\pi}^{D-2}} \frac{\Gamma\kl{\frac{D-1}{2}}}{\Gamma\kl{\frac{D+2}{2}}} \frac{\displaystyle\sum_{n=0}^{\infty} (-s)^n e^{(n+1)\tilde\mu} \frac{z^{\frac{D+3}{2}}}{(n+1)^{\frac{D-1}{2}}} \kl{K_{\nicefrac{(D-1)}{2}}\kl{(n+1)z} + \frac{D}{(n+1)z} K_{\nicefrac{(D+1)}{2}}\kl{(n+1)z}}}{\displaystyle\kl{\sum_{n=0}^{\infty} (-s)^n \frac{e^{(n+1)\tilde\mu}}{(n+1)^{\frac{D-1}{2}}} z^{\frac{D+1}{2}} K_{\nicefrac{(D+1)}{2}}\kl{(n+1)z} }^2} \frac{\lambda_\text{thermal}^{D-1}}{g(D) c_\text{eff} A_\text{H}}$\\
				$\eta_{\text{avg.},\tau,n}$ &$=$& $\displaystyle\frac{D-1}{2 \pi^{\frac{D-2}{2}} z^{\frac{D-1}{2}}} \frac{\Gamma\kl{\frac{D-1}{2}}}{\Gamma\kl{\frac{D}{2}}} \kle{\sum_{n=0}^{\infty}(-s)^n e^{(n+1)\tilde\mu} \kl{\frac{2}{n+1}}^{\frac{D-1}{2}} K_{\nicefrac{(D-1)}{2}}\kl{(n+1)z} }^{-1} \frac{\lambda_\text{thermal}^{D-1}}{g(D) c_\text{eff} A_\text{H}} $\\[0.5cm]
				$\eta_{\text{avg.},\lambda,n}$ & $=$ & $\displaystyle\frac{D-1}{(2z)^{D/2}} \kle{ \sum_{n=0}^{\infty} (-s)^n e^{(n+1)\tilde\mu} \kl{\frac{\pi}{n+1}}^{\frac{D-2}{2}} K_{\nicefrac{D}{2}}\kl{(n+1)z}  }^{-1}\frac{\lambda_\text{thermal}^{D-1}}{g(D) c_\text{eff} A_\text{H}}$
			\end{tabular}
		}
			\caption{Sparsities for massive particle emission in a $D+1$-dimensional Tangherlini space-time in terms of polylogarithms $\Li{n}(x)$, and modified Bessel functions of the second kind $K_\nu(x)$. Here, $z\defi \frac{mc^2}{\kB T_\text{H}}$ is a dimensionless mass-parameter. $\lambda_{\text{thermal}}$ is the thermal wavelength, $c_\text{eff}$ a correction factor to link capture cross-section with horizon area $A_\text{H}$, and $g(D)$ the particles' degeneracy factor.}
			\label{tab:massive}
		\end{table*}
	\section{Conclusion}
	
	In this letter, we have provided a generalisation to $D+1$ dimensions of the exact, heuristic, semi-classical results for non-rotating black holes found in \cite{HawkFlux1} which introduced the concept of sparsity. We have reproduced and improved on the results of \cite{HodNdim}, and shown agreement with previous numerical studies \cite{KantiNdimBH,CardosoBulkHawking,KantiNdimBH2}. This highlights two things: First, it demonstrates the robustness of the heuristic concept of \enquote{sparsity}. Second, this concept provides a quick, simple, and often pedagogical insight into radiation processes, here exhibited on the Hawking radiation from a Tangherlini black hole in $D+1$ space-time dimensions.
	
	Given the propensity of higher dimensional model building encountered in the quest for quantum gravity, it seems important to have an easy-to-calculate, but predictive physical quantity like sparsity that helps to understand differences between such models. This is particularly true for the prime benchmark that is the Hawking effect: Traditionally, a focus for this differentiation between models relies on the connection between entropy and area, and how different models vary this more or less severely compared to the Bekenstein--Hawking result. Using instead a property of the emitted radiation (like sparsity) seems experimentally more readily accessible than entropy or horizon area. Here, we presented the results for models predicting a dynamical situation as higher dimensional general relativity would have. Sparsity is, however, more than just a tool of curved space-time quantum field theory and general relativity: Other phenomenological approaches involving generalised uncertainty principles \cite{SparsityAna,OngGUPSparsity}, and attempts to model backreaction \cite{SparsityBackreaction}, further illustrate the use of this tool also for other dynamics, as more particle physics inspired extensions (like string theory) might imply.
	
	An obvious extension of the present letter is the analysis of Myers--Perry black holes along the lines of the Kerr analysis in \cite{MyThesis} and \cite{HawkFlux1}; for sparsities amenable to the binning property~\eqref{eq:binning} even a combined superradiance-mass analysis could be performed based on the present results. Less straightforward would be an extension to other high-dimensional models not implying dynamics not akin to those of general relativity.
	
	\section*{Acknowledgements}
	Part of the research presented here was funded by a Victoria University of Wellington PhD Scholarship. The author would like to thank Finnian Gray, Alexander Van-Brunt, and Matt Visser for many helpful discussions.

\end{document}